\documentstyle[prl,twocolumn,aps,epsf]{revtex}

\begin{document}

\input{epsf}
\draft

\author{Roland Faller\footnotemark and Lorenz Kramer\\
{\small \it
Physikalisches Institut der Universit\"at Bayreuth, D-95440 Bayreuth, Germany
}
}
\title{Phase chaos in the anisotropic complex Ginzburg-Landau Equation}
\maketitle
\renewcommand{\thefootnote}{\fnsymbol{footnote}}
\footnotetext{$^*$present address: Max-Planck-Institut f\"ur Polymerforschung, 
D-55128 Mainz, Germany}
\begin{abstract}
Of the various interesting solutions found in the two-dimensional complex
Ginzburg-Landau equation for anisotropic systems, the phase-chaotic states show
particularly novel features. They exist in a broader parameter range than in 
the isotropic case, and often even broader than in one dimension. They 
typically represent the global attractor of the system. There exist two 
variants of phase chaos: a quasi-one dimensional and a two-dimensional 
solution. The transition to defect chaos is of intermittent type.
\end{abstract}


The complex Ginzburg-Landau equation (CGLE) plays the role of a generalized
normal form for spatially extended media in the vicinity of a supercritical
Hopf bifurcation involving a non-degenerate (oscillatory) mode. It has a wide 
range of applications extending from hydrodynamic instabilities
\cite{schoepf93,treiber97} and nonlinear optics \cite{nloptics} to oscillatory
chemical instabilities like the Belousov-Zhabotinsky reaction \cite{hynne93} or
oxidation on catalytic surfaces \cite{baer94}. For a general review see e.g
\cite{cross93}.

The one-dimensional (1D) and the 2D isotropic cases have been investigated
rather well \cite{coullet89}-\cite{hecke98}. A number of results have also been
obtained in 3D \cite{aranson97,gabbay97}. Taking up some earlier work
\cite{weber91} we recently reported about spirals and ordered defect chains in
the anisotropic complex Ginzburg-Landau equation(ACGLE) \cite{faller97}
\begin{equation} \label{CGLE}
\partial_tA=[1+(1+ib_1)\partial_x^2+(1+ib_2)\partial_y^2-(1+ic)|A|^2]A.
\end{equation}
Here $A$ is the complex amplitude modulating the critical mode in space and
time. The usual reduced units are used. This equation was also studied in the
context of defect chaos (DC) \cite{roberts96} and wind-driven Eckmann boundary
layers \cite{haeusser97}.

Actually the range of applicability of Eq. (\ref{CGLE}) is considerable. The
isotropic case, i.e. Eq.(\ref{CGLE}) with $b_1=b_2=b$, can essentially be
applied only to isotropic systems undergoing a {\em spatially homogeneous} Hopf
bifurcation. A nonzero wavenumber $q_c$ leads to traveling or standing waves,
as in many hydrodynamic instabilities. Then, in systems that are isotropic in
the basic state, one has a continuous degeneracy of the critical modes, which
makes a more elaborate description necessary. In the presence of an anisotropy,
like e.g. in the well-studied system of electro-convection in liquid crystals
\cite{treiber97}, this degeneracy is typically lifted and Eq.(\ref{CGLE}) is
appropriate. Also, of course, anisotropic systems with a $q_c=0$ bifurcation,
as occur in oscillatory surface reactions \cite{baer94}, require $b_1 \ne b_2$.
Taking linear transformations of $x$ and $y$ into account the term  involving
second derivatives is general. Transforming into a co-moving frame a linear 
group velocity involving a first space derivative vanishes. In the (common)
situation of degeneracy between left- or right-traveling waves, we assume only
one type to survive  (which is often the case).

The ACGLE has a 2D wave-vector band of plane-wave solutions
$A=F\times \exp{i(Qx+Py-\omega t)},\,F^2=1-Q^2-P^2,\,
\omega=c+(b_1-c)Q^2+(b_2-c)P^2$.
They are stable against long-wavelength modulations when
$(1+2\frac{1+c^2}{1+b_1c})Q^2+(1+2\frac{1+c^2}{1+b_2c})P^2 <1$
holds (generalized Eckhaus instability) while the Newell criterion
\begin{equation}
1+b_ic>0,\quad i=1,2
\end{equation}
is satisfied in both directions. From these relations one sees that the stable
$Q$ band shrinks to zero as $1+b_1 c \to 0^+$ (Benjamin-Feir (BF) instability)
with a similar behavior of the $P$ band. Actually, the Eckhaus instability for
$(Q,P) \ne 0$ is of the convective type and plane waves can occur over a 
limited spatial extension in a larger range \cite{aranson92}.

The bifurcation connected with this instability is supercritical when one is at
the BF limit or sufficiently near to it, i.e the amplitude of the destabilizing
side-band modes actually saturates \cite{janiaud92}. However, the resulting 
quasi-periodic solutions, as far as they are themselves modulationally stable,
have for vanishing $(Q,P)$ a small basin of attraction in the BF unstable 
range, and in the studied 1D and isotropic cases the relevant attractors turn
out to be spatio-temporally chaotic. Nevertheless, since the amplitude $|A|$
saturates to a value near 1, only the phase $\Phi$ (we write 
$A=|A| \exp{i\Phi}$) is dynamically active. In 1D the bifurcation at the BF
instability, including slow modulations, is captured by the celebrated 
Kuramoto-Sivashinsky phase equation (see below). It exhibits the so-called
phase chaos (PC) (or phase turbulence).

\begin{figure}[ht]
\begin{center}
\mbox{\epsfxsize 7cm\epsffile{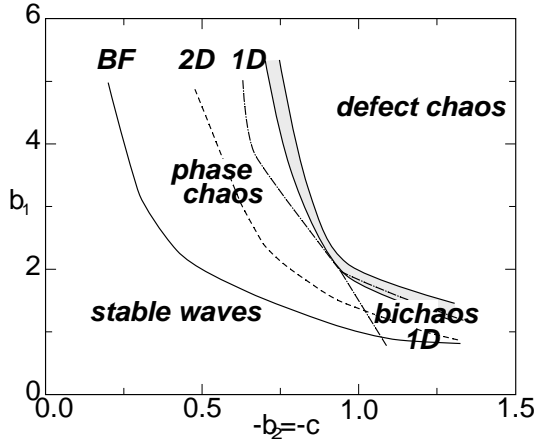}}
\caption{phase diagram for $b_2=c$}
\label{graph}
\end{center}
\end{figure}

PC in the 1D CGLE was studied numerically first by Sakaguchi
\cite{sakaguchi90b}, who also studied the breakdown and crossover to chaos 
involving phase slips (zeros of $A$ in space-time) further away from the BF 
curve. This state is, in analogy to the 2D case (see below), often referred to
as defect turbulence or defect chaos (DC) \cite{coullet89}. The resulting
phase diagram was studied numerically in detail by Shraiman et al.
\cite{shraiman92}, who discovered that for $|b|\ge 1.8$ the transition between
PC and DC is continuous, whereas it is hysteretic with a bi-chaotic region
otherwise (see Fig. \ref{graph}, dashed-dotted line and region marked bichaos
1D). A detailed study with longer simulations and larger systems was performed
by Egolf and Greenside \cite{egolf95}.
A rather exhaustive study in 2D (isotropic case $b_1=b_2$) was presented by
Manneville and Chat\'e \cite{manneville96}. Here the region of PC is somewhat
smaller than in 1D (see Fig. \ref{graph}, broken line). Also, the transition is
always hysteretic, which may be related to the fact that the zeros of $A$ now
correspond to topological defects. The breakdown of PC involves the creation of
pairs of defects of opposite polarity which separate and loose correlation 
("unbind"). Once initiated, the process is self sustaining leading to a nucleus
and eventually to fronts that appear always to invade the PC state 
\cite{manneville96}. Thus in the isotropic case PC is never the globally stable
attractor.

Actually over much of the region where one has PC the global attractor is not 
DC as such, which appears only transiently, but rather a frozen state (vortex
glass) with a disordered distribution of defects 
\cite{huber92,huber95,manneville96}. Every second defect emits a 
spiral wave of the type well known in the Eckhaus-stable range. The emitted 
waves remain intact over finite-sized cells by convective stabilization.
Rotating spirals (time dependence $\propto \exp{i\omega t}$) exist also in the
ACGLE. In spirals the group velocity, which in plane waves is $2(b_1-c)q$ in 
the $x$ direction [$2(b_2-c)p$ in the $y$ direction] is expected to point 
outward in all directions. In order to have coherent wavefronts one needs 
$(b_1-c)(b_2-c)>0$. Our simulations confirm that spirals are found only under
this condition. Also the expected aspect ratio $\sqrt{(b_1-c)/(b_2-c)}$ of the
equiphase lines of spirals is confirmed by the simulations.

Our investigation was motivated in particular by the question of what happens
in the parameter regime $(b_1-c)(b_2-c)<0$ where spirals do not exist, and
therefore also the existence of DC could be questionable. With this inequality
the BF instability (necessary for PC) can only occur in one direction (we
choose $b_1c<-1$, i.e. instability in the x direction) and the anisotropy is
"strong" \cite{weakAni}. Since the ACGLE has the symmetry $(b_1,b_2,c,A) \to
(-b_1,-b_2,-c,A^*)$, we always chose $b_1>0$ (in comparing with other works we
transformed to this convention), and therefore $b_1-c>0$.

The quick answer to the above question is actually quite simple: The system 
remains in PC "longer" than in the isotropic case, but eventually it does 
develop ("anisotropic") DC. Since in DC defects actually hardly emit waves, in
contrast to the situation in the vortex glass, no problem arises with opposite 
group velocities. The investigation led to surprises to be discussed now.

\begin{figure}[ht]
\begin{center}
\mbox{\epsfxsize 4cm \epsffile{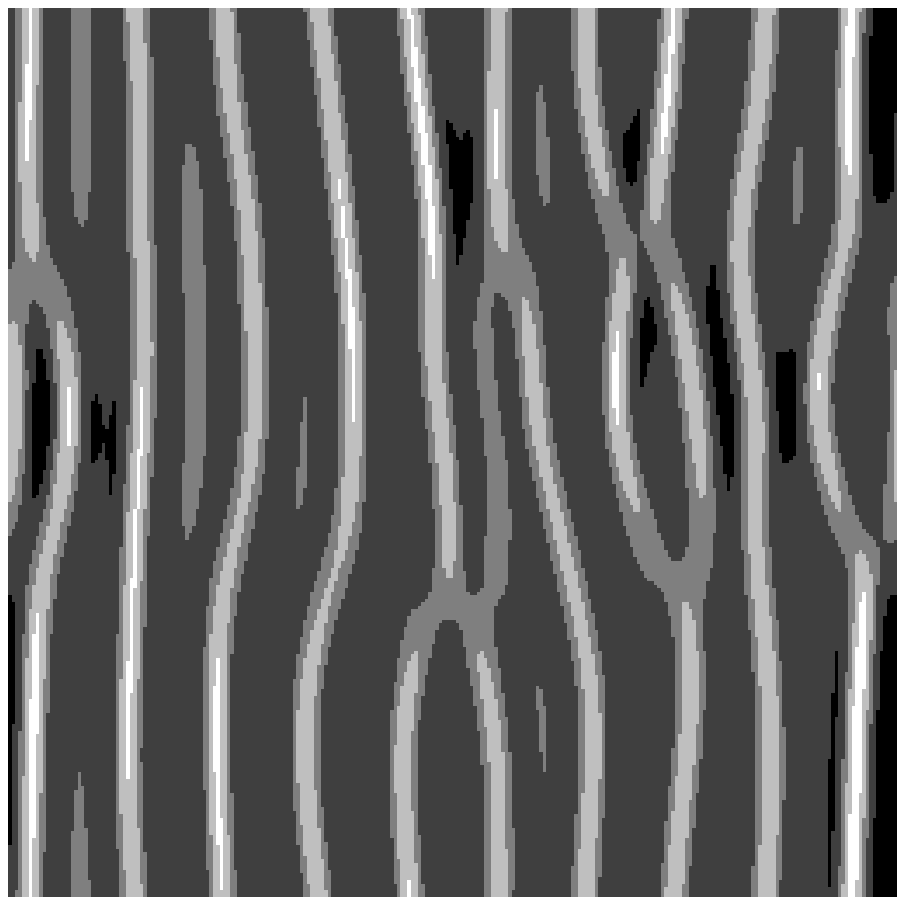}} $\,$
\mbox{\epsfxsize 4cm \epsffile{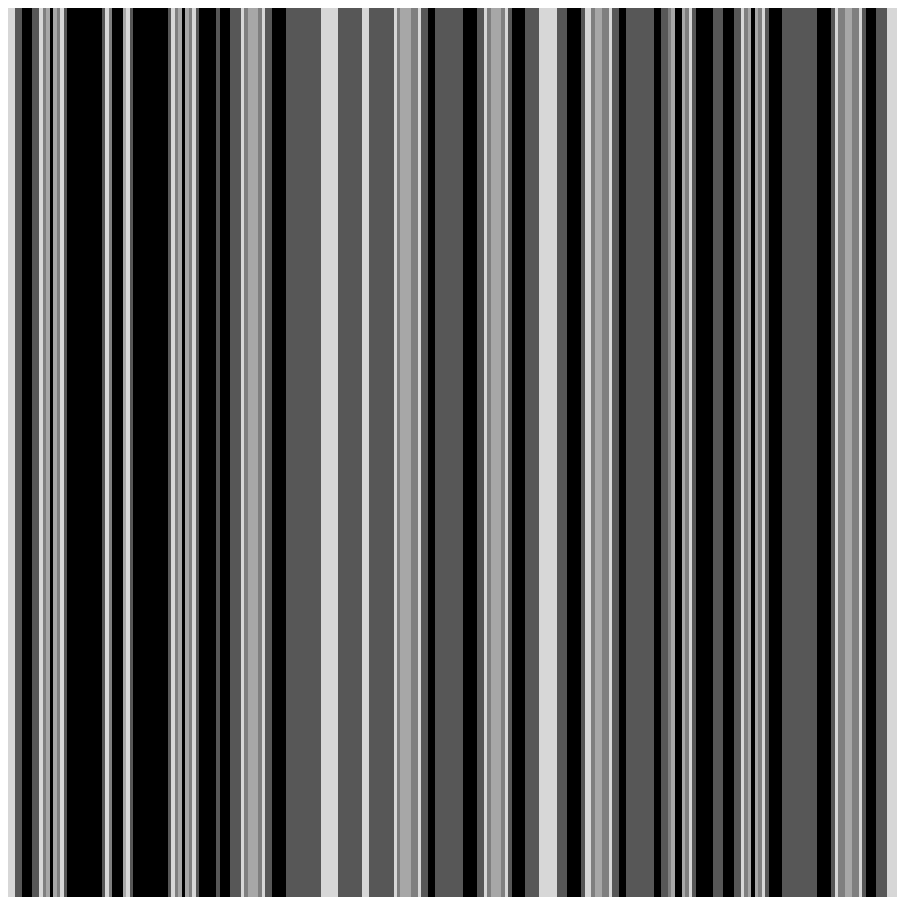}}
\caption{Modulus $|A|$ for $b_1=5.0$, $b_2=c=-0.25$, and system size $L=700$ 
(white: $|A|=1$, black: $|A|\approx 0.9). a): $ 2D PC ("PCII") b): quasi-1D PC 
("PCI").}
\label{fig:pt}
\end{center}
\end{figure}

We have performed detailed simulations of the ACGLE in systems of size $L$ 
between 100 and 2500 dimension-less units with discretization 
$\triangle t \approx 0.1$ and $\triangle x=L/N$ between 0.3 and 5, where $N$ 
is the number of Fourier modes in each direction of the pseudo-spectral 
algorithm used. We used periodic boundary conditions with initial conditions
that imposed a zero phase difference across the system. Hence PC with a nonzero
background wave-vector as studied recently in 1D \cite{gradientPC} was 
excluded. The results depend only weakly on the discretization and on system
size (for sufficiently large systems). Choosing $b_1=b_2$ the results of
\cite{manneville96} could be reproduced. Subsequently we changed $b_2$ in the
direction of $c$. This always increased the range of PC (i.e. $|c|$ could be
chosen larger). The limit of PC for the case $b_2=c$ is depicted in Fig. 
\ref{graph}. To the left of the shaded region no defects were observed, to the
right of it DC was found. The shaded region itself is the parameter range where
we found intermittency (see below). Note that for $b_1 > 2$ even the 1D limit
of PC could be exceeded. Since it turns out that for the effect on PC the sign
of $b_2-c$ is, after all, not decisive, we in fact did many of the studies at
$b_2=c \quad(<0)$. A snapshot of the PC found there is shown in Fig. 
\ref{fig:pt}a. 

We now come to the qualitative new features of anisotropic PC as extracted from
our simulations performed in the range $1\leq b_1 \leq 5$ and $b_2=c$:

(i) PC is the global attractor, i.e. with random initial conditions the system
ends up in PC after the eventual annihilation of transient defects. This is in
contrast to the isotropic case, where PC is never the global attractor.

(ii) In the whole investigated range the transition between PC and DC (as
$b_2=c$ is varied) goes through a stage of intermittency (Fig. \ref{graph}
shaded region), which is not found in the isotropic case. In the intermittent
state defect pairs are created in the form of bursts which subsequently 
annihilate again, keeping the correlation between partners, i.e. defect pairs 
remain bounded. So in this regime, in spite of the presence of defects, phase
coherence persists and the state should therefore be classified as PC. At a 
critical value of $|c|$ ($=c_u$) defects start to unbind rather fast, and this
should be associated with the onset of DC. Recently a transition between two
defect chaotic states in coupled Ginzburg-Landau equations was reported where
one also sees this unbinding of pairs \cite{granzow98}.

\begin{figure}[ht]
\begin{center}
\mbox{\epsfxsize 4.2cm\epsffile{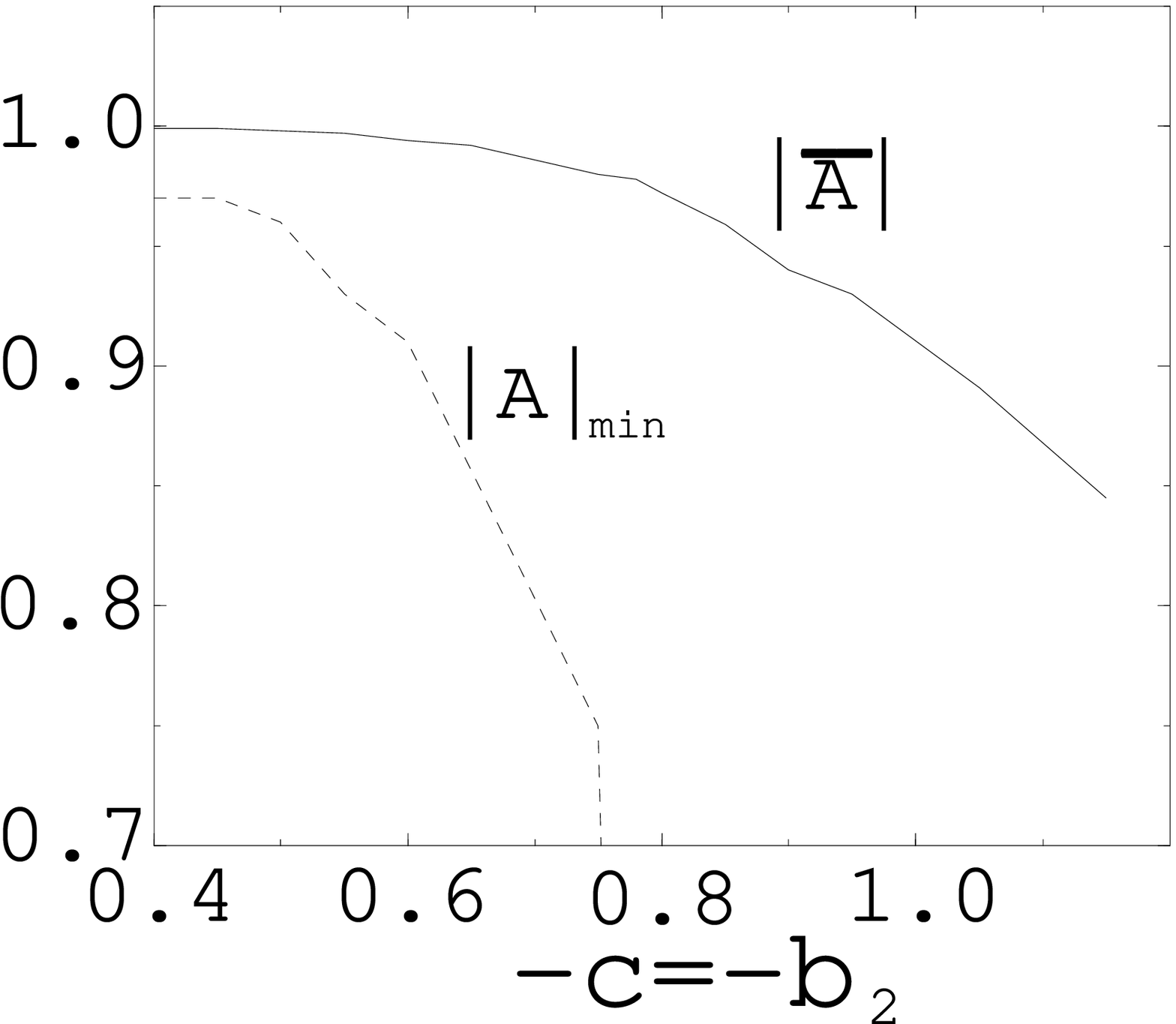}}
\mbox{\epsfxsize 4.2cm\epsffile{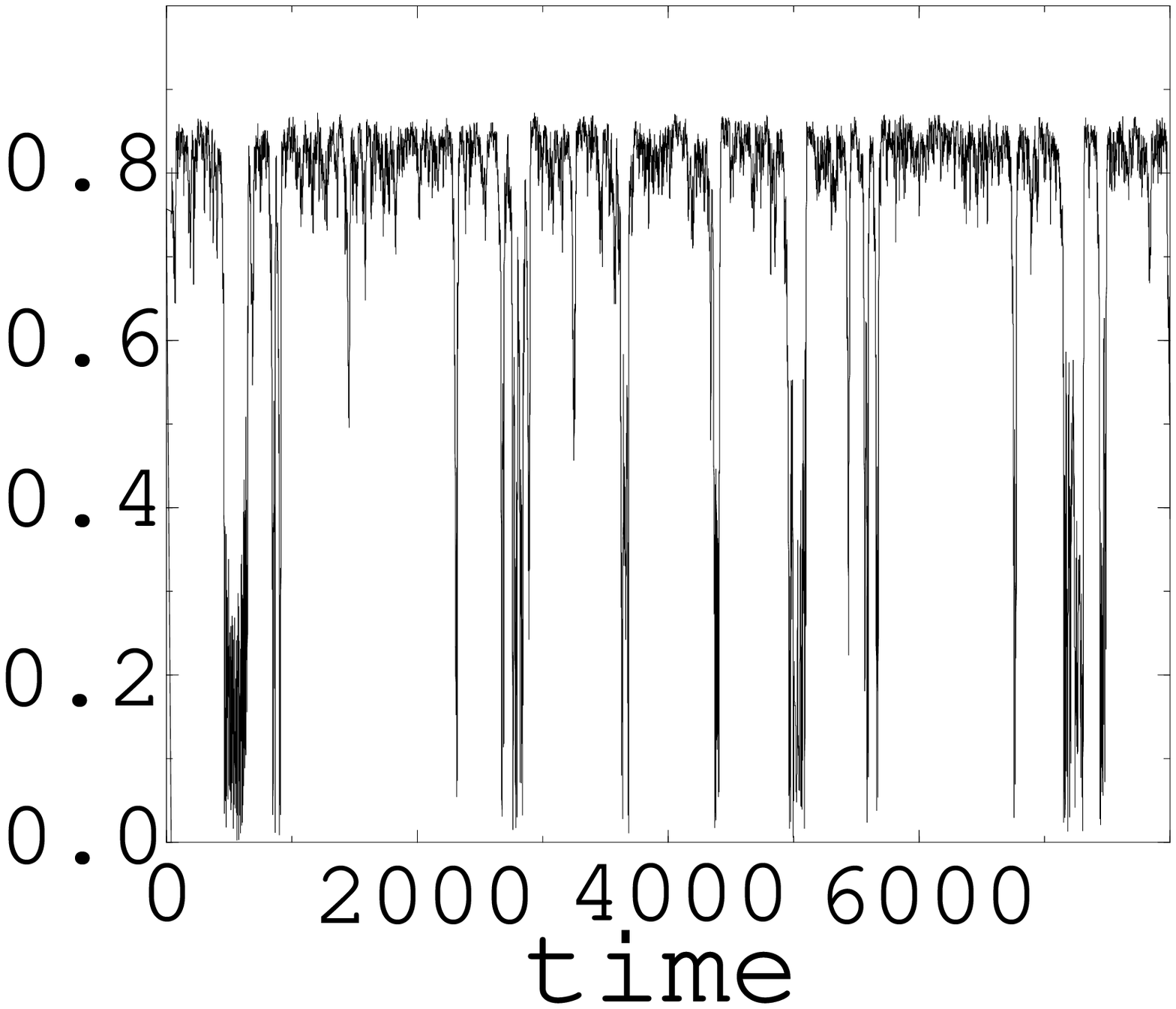}}
\caption{a): $\overline{|A|}$ (solid line) and $|A|_{min}$ (broken line) 
decrease with growing $|b_2|=|c|$ ($L=700,\ b_1=4.0$). b):
time series of $|A|_{min}$ for $b_1=2.0,\,b_2=c=-0.96$}\label{inter}
\end{center}
\end{figure}

In PC the spatial average of the amplitude $\overline{|A|}$ is very close to
$1$, see Fig. \ref{inter}a (solid line). One finds a kink at the onset of
intermittency from where on $\overline{|A|}$ starts to drop faster. The limit
of existence of PC can here be assigned to $|c|$ slightly below $0.8$. Also
shown in Fig. \ref{inter}a is the minimum $|A|_{min}$ of $|A|$ (broken line).
Once $|A|$ falls below $|A|=0.6$ breakthrough to $A=0$ typically occurs. 
Figure \ref{inter}b shows $|A|_{min}$ as a function of time in the intermittent
range.

(iii) In addition to the 2D PC discussed up to now ("PCII") there exists close
to the BF boundary and coexistent with PCII a strictly quasi-1D PC ("PCI") 
with spatial variations only in the unstable $x$ direction (see Fig.
\ref{fig:pt}b for a snapshot). It is obtained by initializing the system with
a function $A$ that differs from $1$ only by small variations in $x$. PCI is
only stable against small perturbations in the $y$ direction and easily
transforms into PCII (it is metastable). At its limit of stability, which for
$b_1=5.0$ is slightly below $|c|=0.3  (=|b_2|)$, the transformation becomes 
spontaneous \cite{remPCI}. Because of CPU time limitations the transition 
could not be studied extensively.

Next we introduce a nonlinear phase equation which should yield a simplified
description of phase chaos becoming exact in the limit $1+b_1c \to 0^-$. Using
the standard procedure \cite{cross93} one arrives at the following equation for
the (strongly) anisotropic situation
\begin{eqnarray}
\partial_t\Phi=-|D_x|\partial_x^2\Phi-D_{x4}\partial_x^4\Phi-
g_{x}(\partial_x\Phi)^2+D_y\partial_y^2\Phi-\label{nlpe}\\\nonumber
g_{y}(\partial_y\Phi)^2-a\big[2\partial_x\Phi\partial_x^3\Phi+(\partial_x^2
\Phi)^2+\frac{2}{b_1} (\partial_x\Phi)^2\partial_x^2\Phi\big], \\\nonumber
D_x=1+b_1c, \quad D_y=1+b_2c, \quad D_{x4}=b_1^2(1+c^2), \\ \nonumber
g_{x}=b_1-c, \quad g_{y}=b_2-c, \quad a=b_1(1+c^2).
\end{eqnarray}
The first three terms on the r.h.s. of Eq. (\ref{nlpe}) make up the 1D
Kuramoto-Sivashinski equation. The higher-order nonlinear terms proportional 
to $a$ were included by Sakaguchi, who showed them to be responsible for the
breakdown of PC, here implied by a blow up of the phase gradient 
\cite{sakaguchi90b}. Actually the last term in square brackets is formally of
higher order than the others, but it could become important for small $b_1$. In
the stable $y$ direction ($D_y>0!$) it suffices to include the two terms shown,
as done by Bar in the equation without the Sakaguchi terms \cite{bar96}. 
Actually, in the parameter range studied by us, the term proportional to 
$g_{y}$ has little influence (for $b_2=c$ it vanishes anyhow).

By rescaling $t,x,y$, and $\Phi$ one can scale the coefficients of the linear
part and the term proportional to $g_x$ to $1$. Introducing the time scale 
$\tau=D_{x4}/|D_x|^2$, the length scales become $l_x=\sqrt{D_{x4}/|D_x|}$ and
$l_y=\sqrt{D_y D_{x4}}/|D_x|$, which is supported by the simulations of PCII. 
Note that when the BF boundary is approached, where $D_x \to 0$, $l_y$ diverges
more rapidly than $l_x$, so that PCII {\it appears} more and more one 
dimensional. Neglecting in Eq. (\ref{nlpe}) the last term in square brackets 
the only relevant parameter is the prefactor of the Sakaguchi terms, which
becomes ${\hat a}=a|D_x|/(D_{x4}g_x)$.

Comparing PCII obtained from simulations of Eq. (\ref{nlpe}) with that of the
ACGLE we find satisfactory agreement except near to the breakdown (for not too
large values of $b_1$). For $b_1=2.0$ and $b_2=c$ we find the breakdown of the
phase description at $b_2=-0.95$ which is in fair agreement with the value 
found for the ACGLE. In a detailed study of the 1D case at $b_1=3.5$ 
\cite{egolf95} the authors found $c_{1d}=-0.75$ in the CGLE and $-0.55$ in the
Sakaguchi equation, whereas we find $c_{2d}=-0.9$ in the ACGLE (with $b_2=c$)
compared to $-0.75$ with Eq. (\ref{nlpe}). PCI is also found in the phase
equation and can at $b_1=5.0$ be maintained stably up to at least $c=-0.28$.
The lowest-order description by Eq. (\ref{nlpe}) with only the first four terms
on the r.h.s. has PCI and PCII as coexisting solutions.

How can one understand the existence of PCI? For a {\it stable} 1D solution, 
i.e. a solution with negative Lyapunov exponents, the (stable) existence of its
quasi-1D analog is clear in the situation of a stable $y$ direction (this is
most easily seen in the phase equation). On the other hand, in PCI one has
positive Lyapunov exponents for fluctuations that vary only in the $x$ 
direction, so there are also positive Lyapunov exponents for sufficiently small
modulation wavenumber $p$ in the $y$ direction. However, this does not
necessarily destroy PCI, since the only condition is that fluctuations flatten
out in $y$, even though they do not decay. We have confirmed by extensive
simulations of PCI at $b_1=5.0,\ b_2=c=-0.26, L=700,$ and $N=256$ that small
perturbations of the form $a_p\exp{ipy}$ with $a_p<0.2$ (at $p \approx 0.1$)
decay asymptotically in a diffusive manner with a phase diffusion constant
around $D_y$. Under the same conditions stochastic perturbations (uncorrelated
on the discretized lattice in real space) decayed up to an amplitude
$a_d<0.01$. Actually, one also expects solutions of Eq. (\ref{CGLE}) of the 
form $A=\exp{(iPy)} B(x,t)$ with phase-chaotic $B$ to exist. Thus PCI 
presumably represents the center of a $P$ band of phase-winding solutions.

Finally we point out that the interpretation of the PCII $\leftrightarrow$ DC 
transition as a vortex binding-unbinding transition probably allows to 
establish PCII as a thermodynamic phase that is qualitatively different from
DC. In PCII, even if a defect pair is created, it remains bounded and 
annihilates again (the unbinding beyond $c_u$ is a cooperative phenomenon).
The question of the conventional forms of PC representing such a state -- in
contrast to being just a (sometimes metastable) variant of DC with a very low
rate of phase slips or defect pair creation -- has indeed stimulated much of
the previous research on PC \cite{shraiman92,egolf95,manneville96}. Actually
also PCI, although it appears to exist only metastably, can presumably be
considered an independent thermodynamic phase because it differs in symmetry.

Clearly much remains to be done. On one hand, finding criteria for the 
occurrence of PCI and methods to calculate the boundary of existence seems a
most interesting problem. On the other hand, a detailed characterization of the
PCII $\leftrightarrow$ DC transition appears desirable.

We have benefitted from discussions with I. Aronson, J. Neubauer, W. Pesch, and
A. Rossberg. Extensive use of high-performance computer facilities at the LRZ
M\"unchen (Cray T90) and the HLRS Stuttgart (NEC SX4), as well as financial
support by DFG (Kr690/4) are gratefully acknowledged.


\begin{thebibliography}{}
\bibitem{weakAni} We have not studied the effect of small anisotropy 
($b_1/b_2\approx 1$) on PT in any detail. On first sight the simulational
results appear similar to those of the isotropic case
\bibitem{remPCI} There is a small jump in $\overline{|A|}$ from about $0.99$
to about $0.98$ (at $b_1=3.5$) at the crossover from PCI to PCII which would
not be seen in Fig. \ref{inter}a.
\end{thebibliography}
\end{document}